\documentclass[preprint]{aastex}
\usepackage{emulateapj5}
\def\stacksymbols #1#2#3#4{\def\theguybelow{#2}
        \def\verticalposition{\lower#3pt}
        \def\spacingwithinsymbol{\baselineskip0pt\lineskip#4pt}
        \mathrel{\mathpalette\intermediary#1}}
\def\intermediary #1#2{\verticalposition\vbox{\spacingwithinsymbol
        \everycr={}\tabskip0pt
        \halign{$\mathsurround0pt#1\hfil##\hfil$\crcr#2\crcr
                \theguybelow\crcr}}}
\def\lta{\stacksymbols{<}{\sim}{2.5}{.2}}

\begin{document}

\title{STELLAR ORBITS AND THE INTERSTELLAR GAS TEMPERATURE IN 
ELLIPTICAL GALAXIES}

\author{William G. Mathews$^1$ \& Fabrizio Brighenti$^{1,2}$}

\affil{$^1$University of California Observatories/Lick Observatory,
Department of Astronomy and Astrophysics,
University of California, Santa Cruz, CA 95064\\
mathews@ucolick.org}

\affil{$^2$Dipartimento di Astronomia,
Universit\`a di Bologna,
via Ranzani 1,
Bologna 40127, Italy\\
brighenti@bo.astro.it}






\vskip .2in

\begin{abstract}
We draw attention to the close relationship between the 
anisotropy parameter $\beta(r)$ 
for stellar orbits in elliptical galaxies 
and the temperature $T(r)$ of the hot interstellar gas. 
For nearly spherical galaxies,
the gas density $\rho$ can be accurately determined
from X-ray observations and the stellar 
luminosity density $\ell_*$
can be accurately found from the optical surface brightness.
The Jeans equation and hydrostatic equilibrium establish
a connection between $\beta(r)$ and $T(r)$
that must be consistent with the observed stellar
velocity dispersion.
Optical observations of the bright elliptical galaxy NGC 4472 
indicate $\beta(r) \lta 0.35$ within the effective radius. 
However, the X-ray gas temperature profile $T(r)$ for NGC 4472 
requires significantly larger anisotropy, 
$\beta \approx 0.6 - 0.7$, 
about twice as large as the optical value. 
This strong preference for radial stellar
orbits must be understood in terms of
the formation history of massive elliptical galaxies.
Conversely, if the smaller, optically determined
$\beta(r)$ is indeed correct, we are led to the important
conclusion that the temperature
profile $T(r)$ of the hot interstellar gas in NGC 4472
must differ from that indicated by X-ray observations,
or that the hot gas is not in hydrostatic equilibrium.
\end{abstract}



\section{Introduction}

The anisotropy of stellar orbits
in elliptical galaxies can be estimated from 
the temperature of the hot
interstellar gas through which they move.
These two quite different galactic attributes are intimately
related by the Jeans equation for the stars and the 
condition for hydrostatic equilibrium in the gas. 
This simple relationship is valuable since 
both the anisotropy and gas temperature 
are difficult to extract from 
optical and X-ray observations, respectively.

Many massive elliptical galaxies are nearly 
spherical (Merritt \& Trembly 1996) and slowly rotating, 
but their stellar velocity ellipsoids are not in general 
isotropic. 
The non-spherical nature of the stellar velocity dispersion 
is represented by the parameter 
$$\beta = 1 - \sigma_t^2 / \sigma_r^2$$
where 
$\sigma_r$ is the radial stellar velocity dispersion 
and $\sigma_t$ is the dispersion in a transverse direction, 
i.e. $\sigma_t^2 = \sigma_{\theta}^2 = \sigma_{\phi}^2$.
If the orbits are predominantly radial, $0 < \beta < 1$, 
the line of sight velocity profile becomes 
more strongly peaked than a Gaussian profile 
with increasing projected radius $R$;  
if the orbits are mostly 
tangential, $-\infty \le \beta < 0$, 
the profile is more flat-topped and becomes broader with 
increasing radius.
Both $\beta(r)$ and the galactic
potential $\Phi(r)$ can be determined
from optical observations of the line of sight
velocity dispersion $\sigma(R)$,
the optical surface brightness distribution
and the deviation of the stellar 
line profiles from a Gaussian 
as expressed by the line-symmetric
coefficient $h_4(R)$ in a Gauss-Hermite expansion
(e,g, van der Marel \& Franx 1993).

Stellar line profiles observed in most massive E galaxies 
indicate a preference for radial orbits with $\beta \sim 0.3$ 
(Bender, Saglia \& Gerhard 1994; Gerhard et al. 1998; 
Saglia et al. 2000).  
Accurate $\beta(r)$ profiles from optical data 
require high quality data
and considerable care in reduction and analysis.
The radially anisotropic nature of 
stellar orbits in 
luminous elliptical galaxies provides important 
and otherwise unavailable information 
about the merger history of these galaxies 
(Naab, Burkert \& Hernquist 1999),   
so an improved or independent determination 
of $\beta(r)$ would be desirable. 

Similarly, the radial variation of the hot gas temperature 
$T(r) \sim T_{vir} \sim 10^7$ K 
in elliptical galaxies depends on the 
spatial and spectral 
resolutions of X-ray detectors and the accuracy 
of the three-dimensional decomposition that converts 
$T$ as a function of projected radius $R$ to physical 
radius $r$. 
In addition, it is often unclear whether the hot gas at any 
radius has a single temperature or a multitude of temperatures
as might be expected if the gas were cooling. 
It is unclear if the gas cools at all. 
For example, 
high resolution X-ray spectra with XMM-Newton of the 
large E galaxy NGC 4636 fail to show emission lines 
expected from gas at intermediate temperatures,  
such as the 0.574 keV OVII line (Xu et al. 2002), suggesting
that the gas is not cooling below $\sim T_{vir}/3$ 
as in classical cooling flows. 
However, Bregman, Miller \& Irwin (2001) detected 
the OVI 1032,1038\AA~ doublet in NGC 4636 
emitted from gas at $T \sim 3 \times 10^5$ K, implying 
that cooling near the expected rate may occur after all. 
As an additional source of confusion, 
the observed X-ray spectra can often be 
significantly improved by assuming two quite different 
discrete temperatures 
at each galactic radius (e.g. Buote 2002 for NGC 1399; 
Buote et al. 2002a and Tamura et al. 2003 for NGC 5044).
While the origin and physical nature of
gas with only two (or a limited continuum of) temperature 
phases may be conceptually problematical, it may 
be related to certain heating mechanisms. 
Nevertheless, there are good reasons for suspecting
the accuracy or interpretation 
of gas temperature profiles $T(r)$ in elliptical galaxies.

In the following discussion 
we illustrate the close relationship between the radial 
variations of $\beta(r)$ and $T(r)$ using data 
for the well-observed giant elliptical NGC 4472. 
While currently available 
data for this galaxy are adequate to illustrate
this relationship, more accurate optical and X-ray 
data will be necessary to verify $\beta(r)$ or $T(r)$ 
with confidence. 

We recognize that NGC 4472 is not perfectly spherical; 
its optical image has been variously classified as E1 or E2, 
However, 
optical determinations of $\beta(r)$ for NGC 4472 
and other nearly-spherical elliptical galaxies 
have been made under the assumption of spherical symmetry
(e.g. Kronawitter et al. 2000). 
Indeed the very concept of the $\beta$ anisotropy 
parameter is only valid in spherical geometry. 
Similarly, the gas temperature $T(r)$ 
determined from X-ray observations of  
NGC 4472 and other nearly spherical elliptical galaxies 
are based on the assumption of spherical symmetry. 
One justification for this is that 
the stellar gravitational potential that confines 
the stars and hot gas within the half-light radius 
is always more spherical than the 
underlying stellar luminosity distribution
(e.g. Brighenti \& Mathews 1996).

Our discussion of the close relationship between 
stellar anisotropy and gas temperature is 
a thinly-veiled request for observers to concentrate 
on those luminous E galaxies for which the most accurate 
X-ray and optical data can be acquired.

\section{Relating $\beta(r)$ and $T(r)$}

The Jeans equation for the radial stellar velocity 
dispersion $\sigma_r$ is 
\begin{equation}
{1 \over \ell_*}{d(\ell_* \sigma_r^2) \over dr} 
+ {2 \beta \sigma_r^2 \over r} = - g
\end{equation}
where $\ell_*$ is the stellar luminosity density 
corresponding to the stellar velocity dispersions
$\sigma_r$ and $\sigma_t$.
The equation for hydrostatic 
equilibrium in the hot interstellar gas is 
\begin{equation}
{1 \over \rho} { d (\rho c^2) \over dr} = - g.
\end{equation}
Here $c^2 = kT / \mu m_p$ is the isothermal sound speed,
$m_p$ is the proton mass and $\mu = 0.61$ is the
molecular weight.
The uniformity of the stellar mass to light ratio 
within the effective radius of NGC 4472 is excellent 
evidence that the hot gas is in hydrostatic 
equilibrium (Brighenti \& Mathews 1997). 
Both equations contain 
the same gravitational term $g = G M(r)/r^2$.
Eliminating this term results in an equation explicitly 
independent of the confining mass: 
\begin{equation}
{d \sigma_r^2 \over d r} + {\sigma_r^2 \over r}
\left[ {d \log \ell_* \over d \log r} + 2 \beta \right]
= {d c^2 \over d r}
+ {c^2 \over r} \left[ {d \log \rho \over d \log r} \right].
\end{equation}
This equation has been discussed previously, particularly in the 
context of the anisotropic orbits of galaxies in rich clusters
that contain hot gas (e.g. Fabricant et al. 1989).
In addition to hydrostatic equilibrium, 
the validity of this equation depends on the assumption that 
the gas pressure exceeds that of any other interstellar 
pressure such as turbulence, cosmic rays, etc. 

If $\beta(r)$ and $T(r)$ 
(as well as $\rho(r)$ and $\ell_*(r)$)
are known from observations, 
then Equation (3) can be solved for 
$\sigma_r(r)$ and the line of sight velocity dispersion 
is then found from 
\begin{equation}
\sigma^2(R) = { \int_R^{r_t}  \sigma_r^2(r)
\left[ 1 - (R/r)^2 \beta \right] 
\ell_*(r) (r^2 - R^2)^{-1/2} r dr 
\over
\int_R^{r_t} \ell_*(r) (r^2 - R^2)^{-1/2} r dr }
\end{equation}
(e.g. Binney \& Mamon 1982).
In this equation $R$ is the projected radius and $r_t$ is some 
radius significantly beyond the limit of the available data
(we assume $r_t = 3 R_e$). 
If $\beta(r)$ is known, 
Equation (3) can be solved for $\sigma_r(r)$ 
for various parameterized temperature profiles $T(r)$ 
until $\sigma(R)$ from Equation (4) agrees with observed velocity 
dispersions.
By this means it is possible to derive the gas temperature 
profile $T(r)$ from a knowledge of $\beta(r)$. 
Conversely, if the gas temperature $T(r)$ is securely 
known from X-ray observations, then 
$\beta(r)$ can be determined by requiring that 
solutions $\sigma^2(R)$ of Equations (3) and (4)
pass through the observations. 
This close relationship between $\beta(r)$ and $T(r)$ 
is possible and valuable because $\ell_*(r)$, $\rho(r)$ and 
$\sigma^2(R)$ are much easier to determine from observations
than either $\beta(r)$ or $T(r)$.

Except for the very central region, 
the stellar brightness in NGC 4472 is well-fit with a 
de Vaucouleurs profile. 
The stellar B-band luminosity density 
$\ell_*(r) \equiv \ell_{*,deV}(r)$ 
is determined by the total luminosity  
$L_B = 7.89 \times 10^{10}$ $L_{B,\odot}$ and the 
effective radius: $R_e = 104'' = 1.733' = 8.57$ kpc, assuming a
distance $d = 17$ Mpc.
However, within a small ``break'' radius, 
$r_b = 2.41'' = 200$ pc, the stellar luminosity flattens to 
$\ell_{*,core}(r) = \ell_{*,deV}(r_b)(r/r_b)^{-0.90}$
(Gebhardt et al. 1996; Faber et al. 1997). 
In Figure 1 we compare the 
(un-normalized) stellar luminosity density $\ell_*(r)$
in NGC 4472 with the hot gas density profile 
$\rho(r)$ determined from X-ray observations.
The remarkable proportionally 
$\ell_*(r) \propto n_e^2(r)$ was first discovered 
by Trinchieri, Fabbiano \& Canizares (1986).

X-ray observations of the hot gas temperature 
in the central region of NGC 4472 are shown in Figure 2.
Only three observations of the gas temperature 
are available in the region within 10 kpc where 
$\beta$ has been determined from optical observations. 
The filled circles show ROSAT deprojected gas temperature 
observations from Buote (2000). 
The linear temperature profile defined by these 
two temperature observations 
\begin{equation}
T(r) = 0.7480 \times 10^7 + 7.251 \times 10^5 r_{kpc}
\end{equation}
is shown in Figure 2 with a solid line.
From Chandra data 
Soldatenkov, Vikhlinin \& Pavlinsky (2003) have recently 
provided an additional temperature measurement of 
$T/10^7~{\rm K} = 0.766 \pm 0.023$ at radius 1'' = 0.0824 kpc
that is shown with a filled square in Figure 2. 
The linear fit to the Buote data passes within the 
error bars of the Soldatenkov et al. observation, 
so all three observations are 
consistent with a linear temperature profile within 
the effective radius $R_e = 8.57$ kpc.
Although the X-ray data is sparse in the region of interest,
it is in perfect qualitative agreement with 
linear gas temperature profiles observed in 
other bright elliptical galaxies in $r \lta R_e$ 
(Fig. 1 of Brighenti \& Mathews 1997).

\subsection{Determining $\beta(r)$ from $T(r)$}

The open circles in Figure 3 show observations 
of the line of sight stellar velocity dispersion 
in NGC 4472 (from Bender et al. 1994) that were used by 
Kronawitter et al. (2000) to determine $\beta(r)$ 
from optical observations alone. 
The filled circles in Figure 3 
are the data of Fried \& Illingworth (1994). 
Both sets of observations lie entirely 
within the effective (half-light) radius, 
$R_e = 104'' = 8.57$ kpc.
For some reason the normalization is systematically different
in the two sets of optical observations.
The Fried-Illingworth data have smaller 
error bars, less overall scatter and  
resemble in overall form the 
velocity dispersion profiles $\sigma(R)$ of other well-observed 
ellipticals [e.g. Gerhard et al. (1998) for NGC 6703]. 
The observational uncertainties apparent in 
the data sets shown in Figure 3 make it difficult 
to determine with confidence any radial dependence 
of the stellar anisotropy.
Consequently, we assume that the orbital anisotropy 
$\beta$ in Equation (3) is constant over the 
range $r > r_o = 0.34$ kpc.

In Figure 3 we compare solutions of Equations (3) and (4) with
the observed line of sight stellar
velocity dispersion in NGC 4472.
Each solution curve $\sigma(R)$ is based on 
the linear gas temperature profile in Figure 2 
and a particular value of $\beta$.
More information about 
the solutions shown in Figure 3 can be found in Table 1.
Of the three solid lines designed to fit the Fried-Illingworth 
data, solution 2 with initial $\sigma_r(r_o) = 430$ km s$^{-1}$ 
and $\beta = 0.7125$ fits the data best.
Solutions 1 and 3 show additional marginally acceptable 
fits although they fall respectively 
below and above the Fried-Illingworth data at smaller 
projected radii $R$.
Evidently $\beta = 0.71 \pm 0.15$ is the most likely 
anisotropy based on the gas temperature observations. 
In some of our solutions of Equation (3) 
$\sigma_r^2 (r)$ becomes unphysically negative at a
radius $r_z$ less than the outer limit of our
integration in Equation (4), $r_t = 3R_e$; 
for these solutions we replace $r_t$ with $r_z$.
In solutions 4 - 6, shown with dashed lines in Figure 3,  
we find that the anisotropy parameter must 
have values $\beta = 0.63 \pm 0.15$
to be in approximate agreement with the data of Bender et al. (1994).
Once $\sigma_r(r_o)$ is selected for 
each of these dashed line solutions in Figure 3, the 
vertical position of the solution 
is extremely sensitive to $\beta$, i.e., 
$\beta$ must be specified to one part in 1000 
to pass through the center of the Bender et al. data. 
One may doubt if Nature can be so discriminating. 
Nevertheless, for either set of optical data, the corresponding 
values of $\beta$ are about twice as large as the 
$\beta(r) \lta 0.34$ found by Kronawitter et al. (2000)
from the Bender et al. data.

\subsection{Determining $T(r)$ from $\beta(r)$}

The preferred orbital anisotropy in NGC 4472 determined by 
Kronawitter et al. (2000) from the Bender et al. observations 
can be accurately represented with 
$$\beta_K(r) = 0.449 r_{kpc}^{0.7333} \exp(-r_{kpc}/2.472),$$ 
which peaks at $r \approx 1.65$ kpc where $\beta_K \approx 0.34$.
In this section we assume that $\beta_K(r)$ is correct 
and solve Equation (3) for $c(r)^2$ and $T(r)$.
As before we consider only linear temperature variations
defined by two temperatures $T_1$ and $T_2$
evaluated respectively at $r_1 = 2.18$ kpc 
and $r_2 = 8.55$ kpc. 
We seek values of $T_1$ and $T_2$ for which solutions 
$\sigma(R)$ of Equation (4) pass through or near the
Bender et al. optical data. 

We begin with a set of solutions for which $\beta = \beta_K(r)$ 
and the temperature profile is constrained to agree with the 
observed linear $T(r)$ in Figure 2. 
When $\ell_*(r)$ and $\rho(r)$ are taken from Figure 1 as before,
the only remaining undetermined parameter required 
to solve Equations (3) and (4) is $\sigma_r(r_o)$.
The four short dashed lines in Figure 4 show 
the line of sight velocity dispersion profile 
$\sigma(R)$ for $269.10 \le \sigma_r(r_o) \le 269.25$ 
km s$^{-1}$, corresponding to solutions 7 - 10 in Table 1.
Since we are using the Kronawitter et al. $\beta_K(r)$, 
it is appropriate to compare these solutions only with the 
data of Bender et al. from which $\beta_K(r)$ was derived.  
Of these, the best-fitting solution is 9, but the overall 
flat or positive slope [$d \sigma(R)/dR > 0$] at larger $R$
does not resemble the $\sigma(R)$ profiles of 
most well-observed E galaxies.
Since none of these solutions is a particularly good 
fit to the Bender et al. data,
it appears that $\beta_K(r)$ found from optical data alone 
is somewhat inconsistent with the 
observed hot gas temperature profile.

To explore alternate X-ray temperature profiles that 
may agree better with the $\sigma(R)$ of Bender et al., we consider
solution 11 in which the hot gas is 
isothermal at the mean value of the two Buote temperatures 
in Figure 2, $T_1 = T_2 = 1.137 \times 10^7$ K.
This solution, shown with a dotted line in Figure 4, 
decreases slightly with $R$, in better agreement with data 
in $R \lta 2$ kpc. 
Solutions with negative $dT/dr$ may fit even better, 
but all known X-ray observations for E galaxies 
have positive $dT/dr$ for $r < R_e$.
Finally, we consider solution 12 based on 
a temperature profile having the same 
slope as the observations in Figure 2, but with uniformly 
higher temperatures, $T_1 = 1.306 \times 10^7$ K and 
$T_2 = 1.768 \times 10^7$ K.
In solution 12 (solid line in Figure 4) the 
radially decreasing line of sight 
dispersion $\sigma(R)$ follows the Bender et al. data 
better than any other curve in Figure 4, 
possibly suggesting that the gas temperature 
should be higher than the observations in Figure 2.
Therefore, either the gas temperatures from X-ray
observations in Figure 2 are too low or $\beta_K(r)$ and the
$\sigma(R)$ data from which it was derived are inaccurate.

\section{Summary and Final Remarks}

We have shown that Equations (3) and (4) establish a useful 
correspondence between the anisotropy of stellar orbits
in elliptical galaxies $\beta(r)$ and the temperature of the 
hot interstellar gas $T(r)$. 
In practice, the $\beta - T$ relation 
is based on the assumption of hydrostatic equilibrium 
for the hot gas and requires reliable observations
of the stellar and gas densities and 
the stellar line of sight velocity dispersions.
This connection between $\beta$ and $T$ is possible 
only if the galaxies are nearly spherical, as emphasized by 
Maggorian \& Ballantyne (2002).

If the gas temperature profile $T(r)$ 
in an E galaxy is securely known 
from X-ray observations, $\beta(r)$ can be 
determined from the observed stellar velocity 
dispersion using Equations (3) and (4).
In principle, this method of determining $\beta(r)$ 
could be done at galactic radii much larger 
than $R_e$ where stellar observations of 
$h_4(R)$ may be difficult but where 
observations of $\sigma(R)$ and  
X-ray gas temperatures are generally reliable.
Conversely, if $\beta(r)$ can be observed with high 
precision from optical observations, the hot gas 
temperature $T(r)$ can be found from Equations (3) 
and (4) by using the gas density 
from X-ray observations.
This means of determining $T(r)$ 
may be particularly valuable in regions 
of the hot gas atmospheres 
where X-ray spectra indicate a two-temperature
or limited multi-temperature thermal structure 
(e.g. Buote 2002b)  
that may complicate the determination of the 
average local gas temperature. 

As a representative example, we have used 
the $\beta$--$T$ relation to determine $\beta(r)$ 
in the bright elliptical galaxy NGC 4472 from 
the hot gas temperature profile 
$T(r)$ found from X-ray observations.
The resulting anisotropy, $\beta \sim 0.71 \pm 0.15$,
indicates that the stellar orbits are considerably 
more radially biased than implied by 
the smaller $\beta_K(r) \lta 0.3$ 
determined from optical observations alone 
(Kronawitter et al. 2000).
These large $\beta$ for NGC 4472 are based on 
the Fried-Illingworth stellar velocity dispersions
$\sigma(R)$.
Somewhat lower anisotropies, $\beta = 0.63 \pm 0.15$ 
are consistent with the velocity dispersion observations of 
Bender et al., but these $\beta$ still exceed the 
$\beta_K(r) \lta 0.3$ found optically by Kronawitter et al. (2000). 
Alternatively, if the Kronawitter et al. 
$\beta = \beta_K(r)$ and the Bender et al. (1994) 
data are correct, then the gas temperature 
in NGC 4472 may be 
higher than previously thought from X-ray observations 
and possibly have a 
different slope $dT/dr$.
These conclusions apply only to the limited region 
$r < R_e$ for which $\sigma(R)$ data currently exist 
for NGC 4472. 
 
It is clear, at least for NGC 4472,
that currently available observations are 
not sufficiently accurate to establish 
either $\beta(r)$ or $T(r)$ with complete confidence.
Accurate optical observations of $\sigma(R)$
and stellar line profiles can be observed with 
8 - 10 meter telescopes to several $R_e$ 
and $T(r)$ is also often known at these large radii.
In addition, it may be possible to determine gas 
temperature profiles with more accurate or prolonged 
X-ray observations than those in Figure 2.
The $\beta(r)$ determined from X-ray emission can serve to 
calibrate the accuracy of purely optical determinations 
or to provide a test for hydrostatic equilibrium.
To accomplish this, it would be desirable to obtain 
high quality stellar dispersion and gas temperature 
observations of those galaxies 
-- such as NGC 4472, NGC 4649, NGC 1399, M87, etc. -- 
that are luminous at both optical and X-ray frequencies.

\vskip.4in
Studies of the evolution of hot gas in elliptical galaxies
at UC Santa Cruz are supported by
NASA grants NAG 5-8409 NAG 5-13275 and NSF grants  
AST-9802994 and AST-0098351 for which we are very grateful.


\clearpage
\vskip.1in
\figcaption[aasbetafig1.ps]{
The observed and azimuthally averaged electron density $n$ 
(cm$^{-3}$) 
profile in NGC 4472 as a function of radius normalized to   
the effective radius $R_e = 8.57$ kpc (distance $d = 17$ Mpc). 
The observations
are from {\it Einstein} (Trinchieri, Fabbiano,
\& Canizares 1986) ({\it filled circles}) and
{\it ROSAT} (Irwin \& Sarazin 1996) ({\it open circles}).
For the inner region
we have Abel-inverted Chandra surface brightness data from
Loewenstein et al. (2001)
({\it open squares}) and normalized them to previous
observations.
The solid line is an analytic fit to the observations.
The dashed line is the square root of the
stellar luminosity density $\ell_*^{1/2}(r)$ with arbitrary 
vertical normalization. 
\label{fig1}}

\vskip.1in
\figcaption[aasbetafig2.ps]{
X-ray observations of the gas temperature in 
NGC 4472 based on deprojected {\it ROSAT} 
observations of Buote (2000) ({\it filled circles}).
Also shown with a filled square 
is the Chandra observation of the gas 
temperature at 0.082 kpc from 
Soldatenkov et al. (2003).
Note that the central temperature is nonzero.
The solid curve is a linear fit for T(r) 
that passes through all three observed temperatures.
\label{fig2}}

\vskip.1in
\figcaption[aasbetafig3.ps]{
The line of sight stellar velocity dispersion 
as a function of projected radius. 
The observations are from 
Bender et al. (1994) ({\it open circles}) and 
Fried \& Illingworth (1994) ({\it filled circles}).
The lines show constant $\beta$ solutions based 
on the 
gas temperature $T(r)$ variation shown in Figure 2.
Each line type corresponds to several  
numbered solutions listed in Table 1 where values 
of $\sigma(r_o = 0.34~{\rm kpc})$ or 
$\sigma(R_e = 8.57~{\rm kpc})$
can be used to 
uniquely identify each line in the plot. 
\label{fig4}}

\vskip.1in
\figcaption[aasbetafig4.ps]{
The line of sight stellar velocity dispersion
as a function of projected radius. 
The data are the same as in Figure 3.
The lines show solution curves based on 
the orbital anisotropy  
$\beta_K(r)$ from Kronawitter et al. (2000).
The numbers associated with each curve 
refer to solutions listed in Table 1. 
Each short dashed line can be uniquely identified with 
a solution in Table 1 by comparing 
values at $\sigma(R_e) = 8.57$ kpc.  
\label{fig5}}

\clearpage

\makeatletter
\def\jnl@aj{AJ}
\ifx\revtex@jnl\jnl@aj\let\tablebreak=\nl\fi
\makeatother

\begin{deluxetable}{rlllccccc}
\scriptsize
\tablewidth{15cm}
\tablecolumns{9}
\tablecaption{SOLUTIONS OF EQUATIONS (3) AND (4)}
\tablehead{
\colhead{Solution} &
\colhead{$\beta$\tablenotemark{a}} &
\colhead{$T_1$\tablenotemark{b}} &
\colhead{$T_2$\tablenotemark{b}} &
\colhead{$\sigma_r(r_o)$} &
\colhead{$\sigma_r(R_e)$} &
\colhead{$r_z$\tablenotemark{c}} &
\colhead{$\sigma(r_o)$} &
\colhead{$\sigma(R_e)$} \\
\colhead{} &
\colhead{} &
\colhead{($10^7~{\rm K}$)} &
\colhead{($10^7~{\rm K}$)} &
\colhead{(km s$^{-1}$)} &
\colhead{(km s$^{-1}$)} &
\colhead{(kpc)} &
\colhead{(km s$^{-1}$)} &
\colhead{(km s$^{-1}$)} \\
}
\startdata
$T \Rightarrow \beta$ & & & & & & & & \cr
1  & 0.6333   & .906 & 1.368 & 390.  & 249. & 25.6   & 282. & 172. \cr
2  & 0.7125   & .906 & 1.368 & 430.  & 253. & 23.7   & 293. & 162. \cr
3  & 0.7775   & .906 & 1.368 & 470.  & 267. & $>3R_e$& 306. & 171. \cr
4  & 0.4766   & .906 & 1.368 & 330.  & 276. & $>3R_e$& 306. & 295. \cr
5  & 0.63     & .906 & 1.368 & 390.  & 290. & $>3R_e$& 318. & 270. \cr
6  & 0.773    & .906 & 1.368 & 470.  & 326. & $>3R_e$& 334. & 245. \cr
  &  &  &  &  &  &  &  &  \cr
$\beta \Rightarrow T$ & & & & & & & & \cr
7  &$\beta_K$& .906 & 1.368   & 269.10& 205.& $>3R_e$& 247. & 210. \cr
8  &$\beta_K$& .906 & 1.368   & 269.15& 219.& $>3R_e$& 272. & 281. \cr
9  &$\beta_K$& .906 & 1.368   & 269.20& 232.& $>3R_e$& 295. & 337. \cr
10 &$\beta_K$& .906 & 1.368   & 269.25& 244.& $>3R_e$& 317. & 385. \cr
11 &$\beta_K$&1.137 & 1.137   & 330.38& 254.& $>3R_e$& 315. & 303. \cr
12 &$\beta_K$&1.306 & 1.768   & 332.9 & 258.& $>3R_e$& 316. & 295. \cr
\tablenotetext{a}{$\beta_K$ refers to $\beta(r)$ from 
Kronawitter et al. (2000).}
\tablenotetext{b}{The linear temperature variation is defined 
by values $T_1$ at $r_1 = 2.18$ and $T_2$ at $r_2 = 8.55$ kpc.}
\tablenotetext{c}{The radial velocity dispersion $\sigma_r(r)$
goes to zero at $r_z$.}
\enddata
\end{deluxetable}

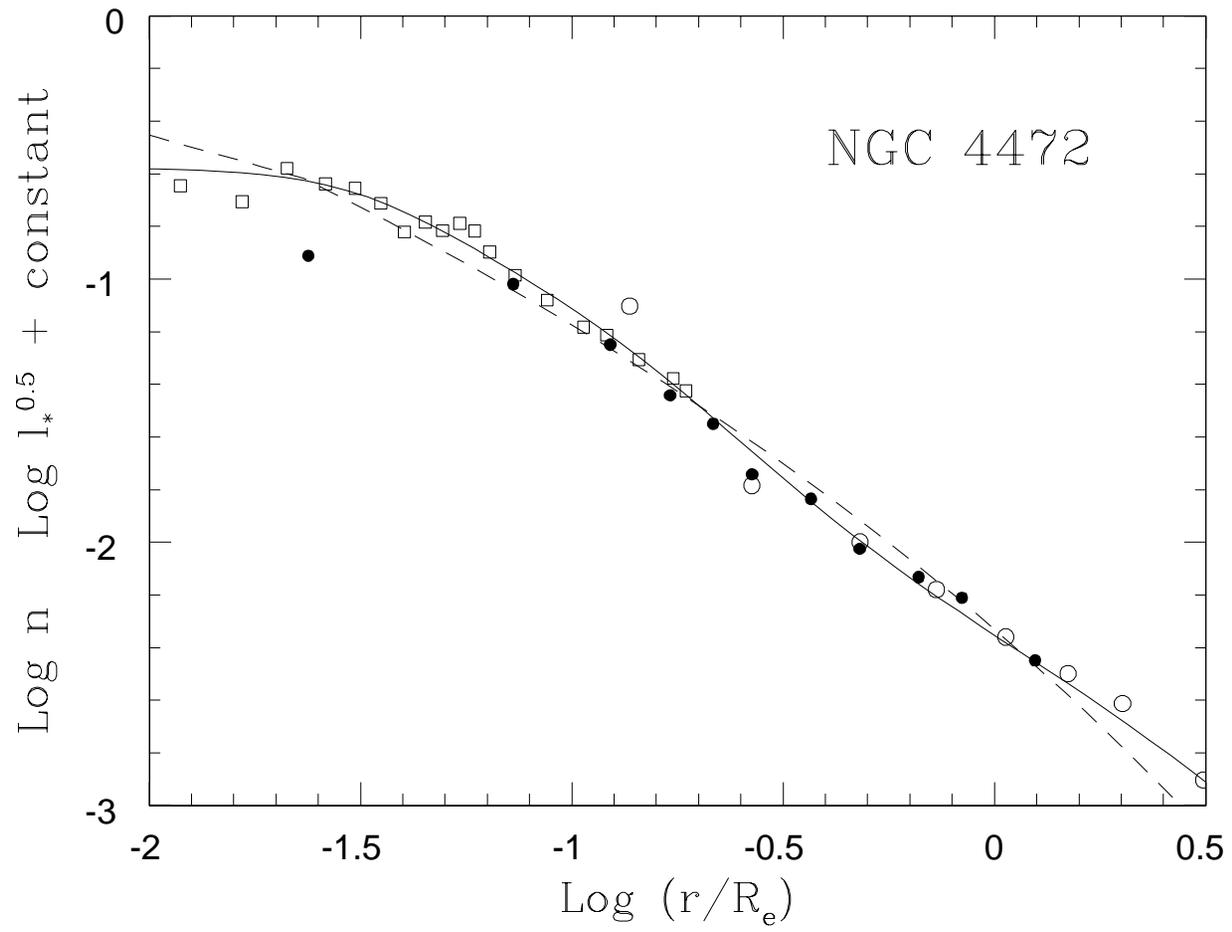
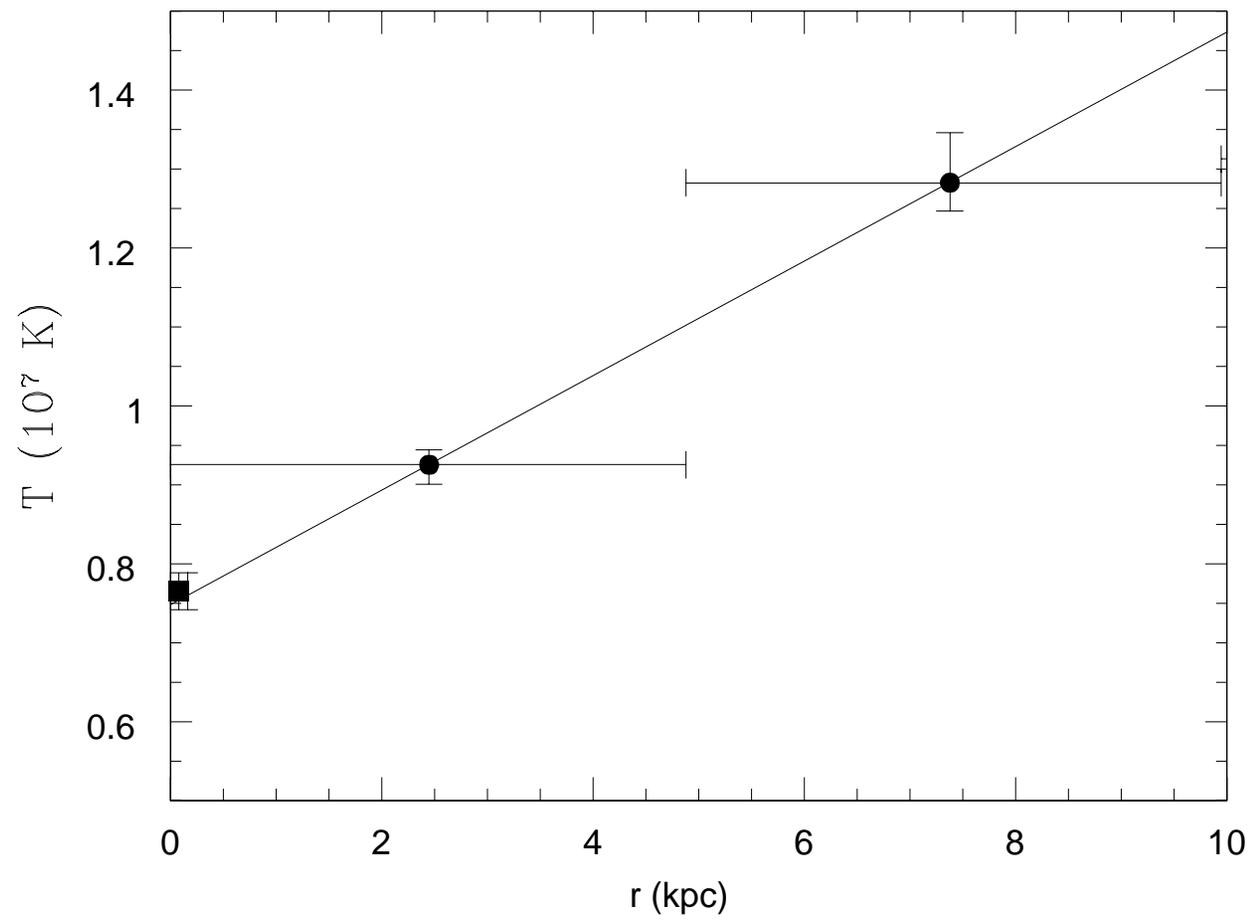
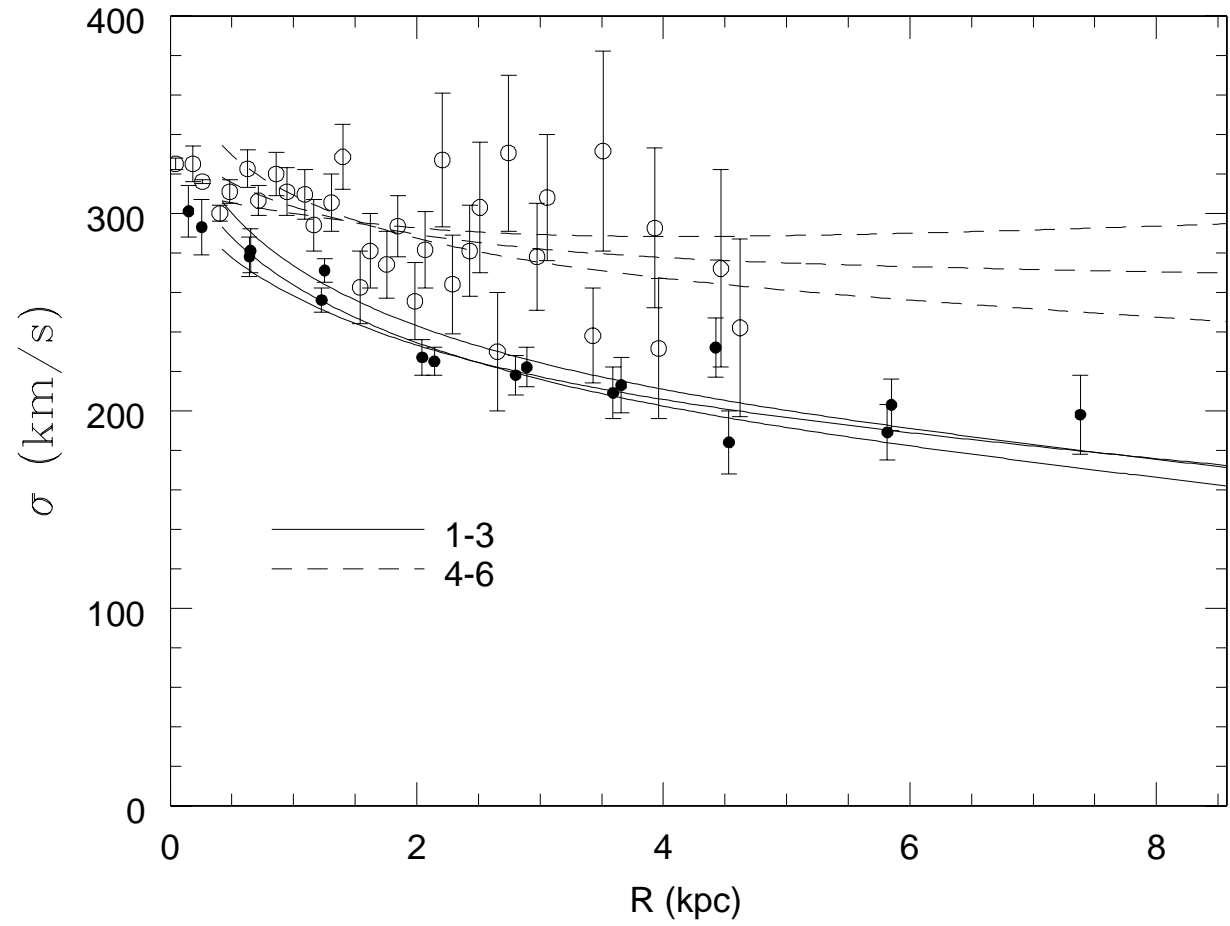
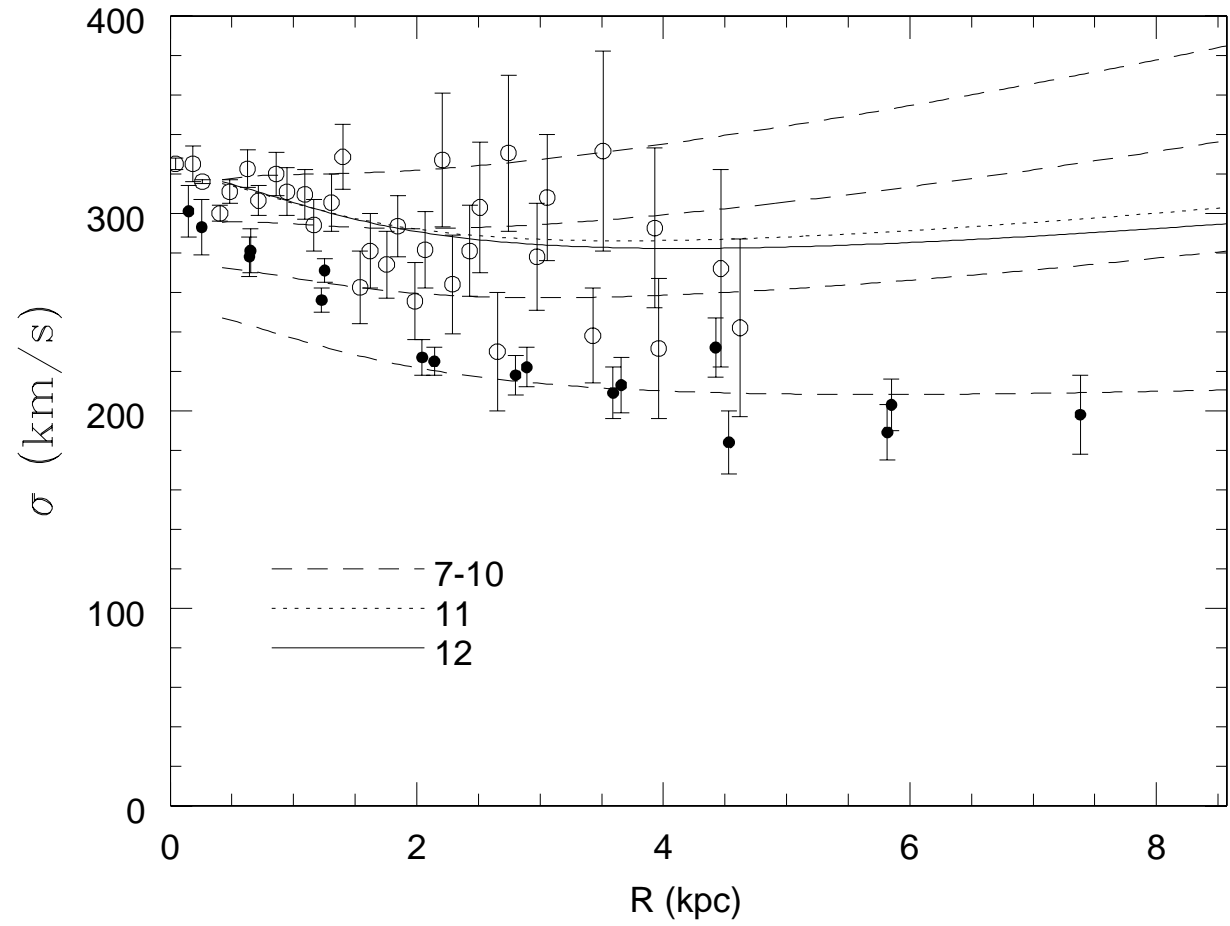
\end{document}